\documentclass[12pt,preprint]{aastex}

\shorttitle{YSO Near-Infrared Emission Morphology}
\shortauthors{Tannirkulam et al.}

\begin{document}




\title{ Strong Near-Infrared Emission  Interior  to the Dust-Sublimation  Radius of Young Stellar Objects MWC275 and AB~Aur.  }



\author{A.~Tannirkulam\altaffilmark{1},
J.~D.~Monnier\altaffilmark{1}, R. Millan-Gabet\altaffilmark{2} , T.~J.~Harries\altaffilmark{3}, E. Pedretti\altaffilmark{4}, T.~A.~ten Brummelaar\altaffilmark{5}, H. McAlister\altaffilmark{5},
N. Turner\altaffilmark{5}, J. Sturmann\altaffilmark{5}, L. Sturmann\altaffilmark{5}}
\altaffiltext{1}{atannirk@umich.edu: University of Michigan, Astronomy Dept, 
500 Church Street, 1017 Dennison Bldg, Ann Arbor, MI 48109-1042}
\altaffiltext{2}{Michelson Science Center, Pasadena, CA}
\altaffiltext{3}{University of Exeter, School of Physics, Stocker Road, Exeter, EX4 4QL}
  \altaffiltext{4}{University of St. Andrews, Scotland, UK}
\altaffiltext{5}{CHARA, Georgia State University, Atlanta, GA}


\begin{abstract}
\noindent 
Using the longest optical-interferometeric baselines currently available, we have detected strong near-infrared (NIR) emission from inside the dust-destruction radius of Herbig Ae stars MWC275 and AB~Aur. Our sub-milli-arcsecond resolution observations unambiguously place the emission between  the dust-destruction radius and the magnetospheric co-rotation radius.  We argue that this new component corresponds to hot gas inside the dust-sublimation radius, confirming recent claims based on spectrally-resolved interferometry and dust evaporation front modeling. 
\end{abstract}




\section{Introduction}
\label{intro} 
\noindent
The inner tenths of an AU in disks around  young stellar objects (YSOs)  hosts   fascinating astrophysical phenomena like disk winds, jets, dust sublimation/condensation and thermal annealing, and gas in-fall onto the star. To make spatially resolved observations of these phenomena, milli-arcsecond angular  resolution is needed  even for nearest the systems. Using  the sub-milli-arcsecond resolution capabilities of the CHARA \citep{ten2005} interferometer array (operated by Georgia State University on Mt.Wilson), we probed  inner disks of the YSOs MWC275 and AB~Aur at 0.1AU scales and found NIR emission  within the dust destruction radius in these systems.

MWC275 (A1e, d=122pc, V=6.86, K=4.59, M$_{*}$=2.3M$_{\odot}$) and AB~Aur (A0pe, d=144pc, V=7.01, K=4.37, M$_{*}$=2.4M$_{\odot}$) are  Herbig Ae stars - pre-main-sequence stars of  intermediate mass. This class of stars shows a significant excess of near-infrared (NIR) flux \citep{natta2001} over the stellar photospheric emission. Current theories \citep{dullemond2001, isella2005, Tannirk} explain this excess solely on the basis of a ``puffed up inner dust rim''. In these ``rim'' theories, dust sublimation terminates the circumstellar dust disk  at a finite radius (the truncation radius depends on the luminosity of the central star and dust properties) creating an optically-thin cavity \citep{rmg1999, lkha2001, eisner2004} around the star. The outer edge of the cavity forms a  dusty rim that traps stellar photons, re-radiating the energy mostly in the NIR. The rim idea seems to broadly explain much of the correlation between the NIR sizes and luminosities of the stars observed in a number of YSO systems \citep{monnier2002a, monnier2005, rmg2007}.

However, recent results have begun to indicate that  models in which the entire NIR excess arises solely  from dust rims are inadequate  in explaining some  of the observed YSO properties. \citet{akeson2005a} noted that  NIR emission from the  the dust rim alone was not sufficient  to explain the spectral energy distribution (SED) of some TTauri stars. \citet{isella2006} pointed out that the dust evaporation temperatures of well studied  astronomical dust types \citep{pollack}  were too low (producing dust rims with large inner radii) to explain the  anomalously small NIR size of AB~Aur.  \citet{eisner2007} and  \citet{eisner2007b} had to invoke gas emission to account for the apparent decrease in NIR sizes of Herbig Ae/Be stars with decreasing wavelength. Most recently, \citet {kraus2007} showed that the NIR visibilities of Herbig Be star MWC147 could not be explained without including emission from inside the dust evaporation radius. Due to the lack of sufficient spatial resolution, some of these conclusions are tentative and based on indirect arguments with multiple interpretations.

\section{Observations}
To unambiguously  constrain NIR disk structure in YSOs, we began an observation campaign with the CHARA Array using the ``CHARA Classic'' beam combiner \citep{ten2005} . The targets  were observed at K band (central wavelength of 2.13 microns) with the long baselines of CHARA at a variety of orientations in 8  runs between June 2004 and June 2007. The longest baseline observation for MWC275 was 325m (resolution \footnote {Resolution is defined as $\frac{\lambda}{2D}$ -lambda is wavelength of observation and D is the interferometer baseline length.}  of 0.67 milli-arcseconds) and 320m (resolution of 0.68 milli-arcseconds) for AB~Aur. The data were reduced using standard CHARA reduction software \citep{ten2005} and these results were cross checked with an independent code developed at University of Michigan (the full calibrated OIFITS data set will be posted on OLBIN - http://olbin.jpl.nasa.gov). HD164031 , HD166295 \citep{merand}  and HD156365 with uniform-disk (UD) diameters of 0.83$\pm$ 0.08 mas, 1.274$\pm$0.018 mas and 0.44$\pm$0.06 mas  were used as calibrators for MWC275. AB~Aur visibilities were calibrated with HD29645 (UD diameter=0.54$\pm$0.07 mas) and HD31233 (UD diameter=0.76$\pm$0.13 mas). The visibility errors (which include calibration errors, statistics and errors due to finite calibrator size) are at the $\sim$6\% level, typical for CHARA Classic. The new fringe-visibility data from CHARA  were combined with  past measurements from IOTA \citep{monnier2006}, PTI \citep{eisner2004} and the Keck Interferometer \citep{monnier2005}   providing a baseline coverage from 15m-320m on the targets, placing strong constraints on the geometry of the inner-most disk regions.\\ 

The baseline orientations used have allowed us to clearly detect the asymmetry of the MWC275 disk (details in a followup paper), as having inclination=48$^{o}{\pm}$2$^{o}$, PA=136$^{o}{\pm}$2$^{o}$, consistent with the inclination of 51$^{+11}_{-9}$ degrees, and PA of 139$^{o}{\pm}$15$^{o}$ determined in \citet{wassell2006}. Thus, in Fig \S\ref{vis_curve} the visibility of MWC275 is shown as a function of ``effective baseline'' - $$B_{eff} = B_{projected} \sqrt{cos^2(\theta) + cos^2(\phi)sin^2(\theta)}  $$ where $\theta$ is the angle between the uv vector for the observation and the major axis of the disk and $\phi$ is the inclination of the disk.  Under the flat disk assumption, the effective baseline  correctly accounts for the change in resolution due to the disk inclination and PA (geometry of finitely thick disks is represented only approximately with  optical depth effects and 3-D geometry of thick disks not being taken into account), allowing us to plot the visibility measurements as a function of one coordinate, simplifying presentation and analysis. As the AB~Aur disk is near face on \citep{eisner2004},  we have plotted its visibility as a function of true projected baseline (Fig \S\ref{vis_curve}).
\label{observe} 

\section{Modeling}
\begin{figure}[h]
\begin{center}
{
\includegraphics[angle=90,width=3.in]{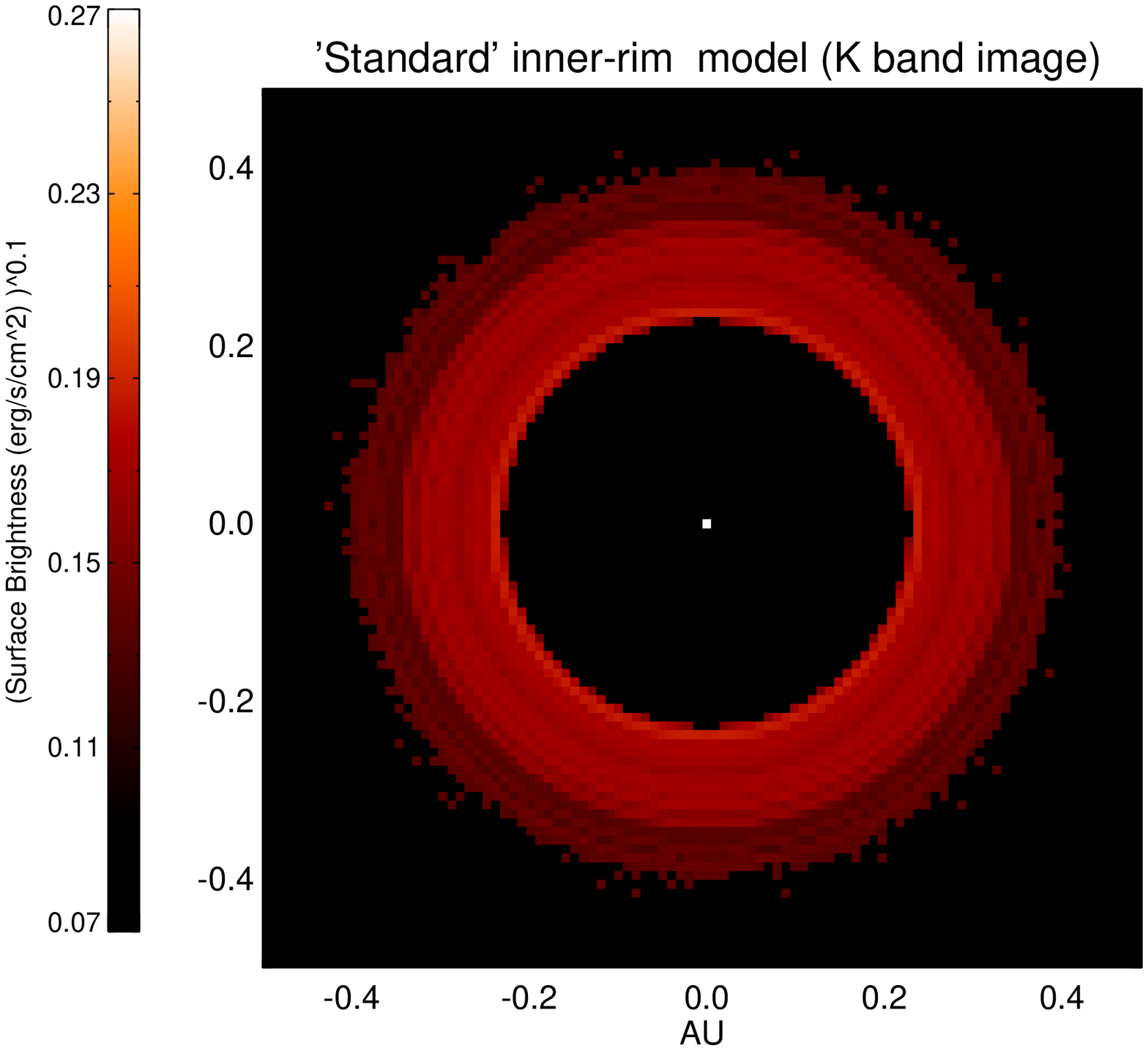}
\includegraphics[angle=90,width=3.in]{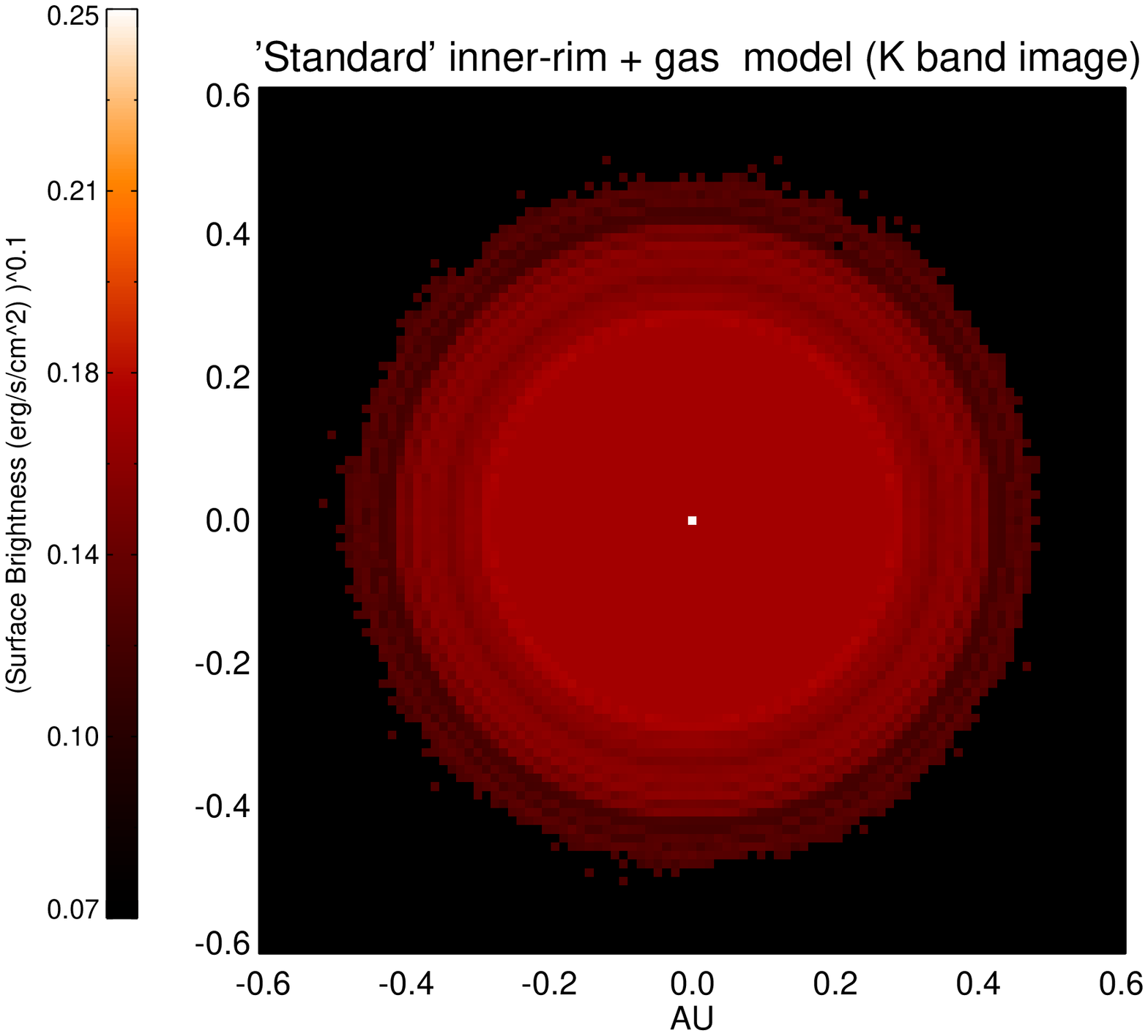}
}

\caption{
 Representative models for near-IR emission in Herbig Ae stars. a) Left panel. A  standard curved dust-rim-only model \citep {isella2005, Tannirk} where the dust sublimation temperature is a power-law function of gas density \citep{pollack}. b) Right panel. Gas emission (modeled as a uniform disk centered on the star)  has been added inside  the dust rim in order to smooth out the emission profile.
}
\label{rim_atm}
\end{center}
\end{figure}

To elucidate the structure of the inner disk and the dust rim, we constructed  dust-only rim models for MWC275 and AB~Aur and fit the models to the visibility data. In these models, all of the K-band emission is assumed to come  from the inner dust rim and the central star (``standard'' model). The rim is in hydrostatic equilibrium and its shape is set by the pressure dependence of the dust grain evaporation temperatures \citep{isella2005, Tannirk}. The normalization of the pressure dependence of the evaporation temperatures (discussed in more detail in a followup paper)  is adjusted so that the model rims fit the short baseline ($<$ 100m) visibility data. As can be seen in Fig \S\ref{rim_atm}, a realistic treatment of dust physics produces a curved inner rim with a sharp inner edge, set by dust destruction,  and a sharp outer edge,  set by the fact that the rim shadows \citep{dullemond2001} a portion of the disk behind it preventing direct star-light from reaching the region.

We are able to  fit  the standard model  to the visibility data  for MWC275 and AB~Aur at baselines shorter than 100m (Fig \S\ref{vis_curve}) fairly well \footnote {The near-IR disk size obtained from the
Keck Interferometer data is $\sim$20\% larger than the size obtained with the CHARA
data (June2004-Aug2006 epochs). The size determined from the S2W1 June2007 data
also differs at the $\sim$25\% level  from the size obtained from earlier CHARA
epochs. The size fluctuations could be related to the observed variability in the infrared photometry of MWC275 \citep{sitko2007}. Variability does not affect any of our conclusions.} (a result already established in multiple previous studies; \citet{monnier2002a, eisner2004, isella2006}~). Rim radii from the model fit are  0.19 AU (thickness 0.04 AU) for MWC275 and 0.21AU (thickness 0.05 AU) for AB~Aur (see Table \S\ref{modelprops} for details). However, the standard model fails glaringly at fitting the long-baseline data. For  baselines longer than 110m the standard rim predicts  strong bounces in the visibility response which are not observed in the data.

The lack of bounce in visibility cannot be explained  in the framework of the standard rim model. Models for  NIR excess in Herbig Ae stars in which all of the excess arises in rims will display bounces in visibility at long baselines because of the presence of sharp ring-like features with high spatial frequency components in the corresponding images, even for the smoothest rims physically plausible.

The only explanation for the absence of long-baseline visibility bounce in the data is the presence of an additional smooth emission component  within the dust destruction radius that suppresses the large visibility bounces seen in a rim-only model (the smooth emission component has to be on size scales similar to the dust sublimation radius and cannot arise from a compact source like magnetospheric accretion). Fig \S\ref{rim_atm} shows a standard model to which   NIR emission has been added within the dust evaporation radius.  Our data cannot uniquely constrain  this emission profile and we assume a constant surface brightness disk (a uniform disk), which is a  simple model. A rim + uniform disk model in which the uniform disk  emission (refer Table \S\ref{modelprops}  for details) accounts for  40-60\% of the total K-band emission provides a good fit to the visibility (Fig \S\ref{vis_curve}) over the whole baseline range. 

Besides the interferometry, emission inside the dust evaporation radius can also help explain the NIR flux  budget of MWC275 and AB~Aur. MWC275 and AB~Aur have  total  K-band emission a factor of 10 larger than the  stellar photospheric K-band emission, whereas the total emission produced by standard rim models \citep{isella2005} is only a factor of 3.5 larger  (SED analysis will be presented in a followup paper) than the photospheric emission. The emission component inside the dust rim  can supply  the  ``missing'' K-band flux in MWC275 and AB~Aur. In  the next section we discuss candidates for the physical origin  of this emission.

\section{Physical Interpretation}
We consider 3 opacity candidates for providing the smooth K-band emission component: a) a  dusty spherical halo around the stars \citep{vinko2006}, b)  highly optically transparent/refractory dust  extending inside the rim and c) gas inside the dust evaporation front. \citet{vinko2006} showed that halo models can fit NIR SED and some early NIR interferometry data on Herbig Ae stars. However, the detection of asymmetries in NIR emission of MWC275 (Section \S\ref{observe}) and other Herbig Ae/Be stars \citep{eisner2004} support a disk model over the spherically symmetric halo model of \citet{vinko2006}. The gas densities in the inner disk are not high enough to let dust survive beyond 1800K \citep{pollack}, making optically transparent/refractory dust an unlikely candidate for the excess K-band emission.

Hot gas is an excellent  candidate as a NIR emission source inside the dust sublimation radius. Molecular gas is known  to have dense NIR opacity features at kilo-kelvin temperatures \citep{carr2004, ferguson2005}.  The required gas emission  levels to explain the long-baseline visibility data can be obtained with optically thin gas emission (optical depth $\tau\sim$ 0.15) with a temperature range of 2500K-3500K \citep{muz2004, eisner2007b}. Our claim regarding the presence of gas emission inside the dust rim  is further supported by the detection  of the  Br$\gamma$ line  in young star HD104327 \citep{tatulli} on size scales similar to the dust destruction radius. 

The required color excess (E(H-K)) from gas emission to match the SED in MWC275 and AB~Aur is  $\sim$ 0.1. MWC275 and AB~Aur have Br$\gamma$ emission equivalent widths of $\sim$6\AA~\citep{lopez2007}. The equivalent widths are in principle sufficient to explain  the color excess with free-free emission  based on the observed correlation between color excess and Br$\gamma$ equivalent width for Be stars \citep{Howells2001}. Thus free-free emission is also a good candidate for NIR emission inside the dust evaporation front. 

\begin{figure}[h]
\begin{center}
{
\includegraphics[angle=90,width=3in]{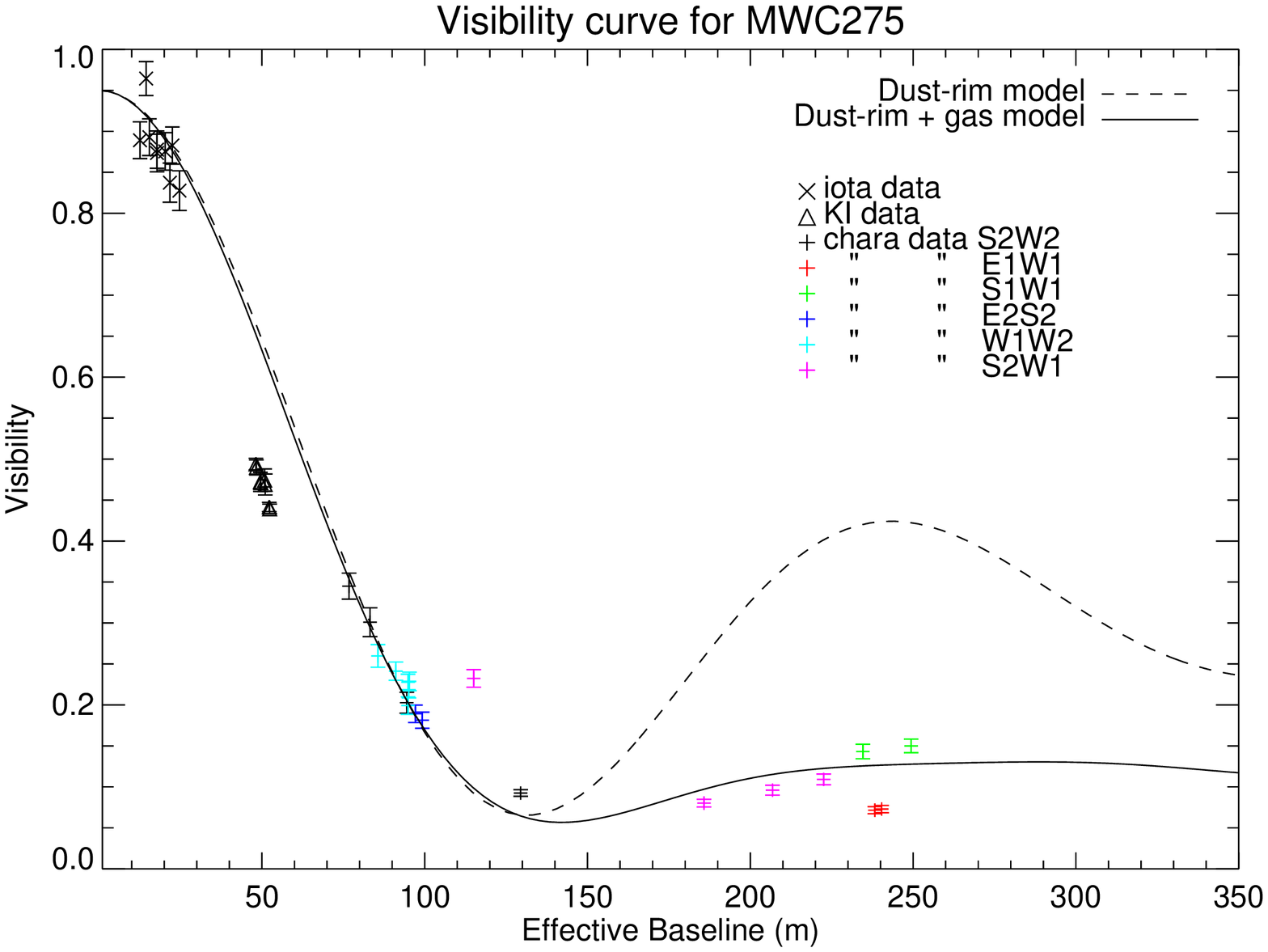}
\includegraphics[angle=90,width=3in]{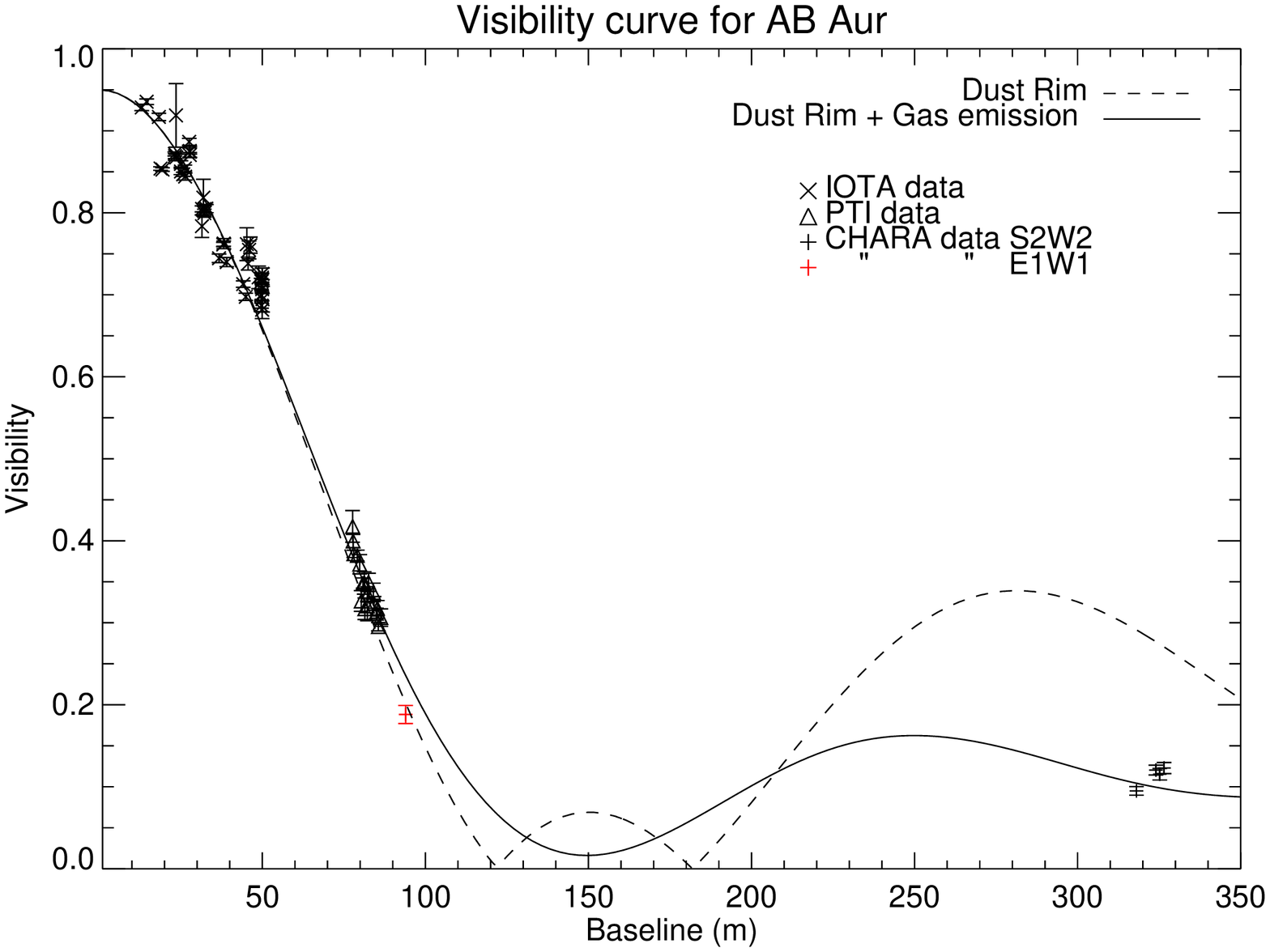}
}

\caption{
  Visibility data and models for  MWC275 (left) and AB~Aur (right). MWC275  
visibilities are plotted as a function of ``effective baseline'' (see section \S\ref{observe} ), which accounts for the change in resolution due to  disk inclination and PA.  \newline The dotted lines correspond to  ``Standard'' rim models tuned to fit visibility data for each source for baselines shorter than 100m.  The solid lines correspond to dust rim+gas models. The model parameters are listed in Table \S\ref{modelprops}.
 }
\label{vis_curve}
\end{center}
\end{figure}

 At first thought, presence of binary companions or source variability would also seem good explanations for the excess K-band emission and the lack of visibility bounce at long baselines. However, the presence of a binary companion within 0.5 AU of MWC275 that contributes $\sim$50\% of the total  near-IR emission will cause changes of $\sim$0.5 in the absolute visibility of the system at any given baseline over the multiple CHARA observation epochs. This variation is not seen in the data (see Fig. \S\ref{vis_curve}). Furthermore, \citet{monnier2006} obtained NIR visibility and closure-phase data on MWC275 and found no evidence for a binary companion  ruling out binarity as an explanation for the observations. A comparison between the Keck Interferometer and CHARA data (taken in different observation epochs) in Fig. \S\ref{vis_curve} shows that MWC275 visibilities are time variable at the $\sim$20\% level, suggesting that variability could complicate our analysis. However,  the data from the S2W1 CHARA telescope pair (Fig. \S\ref{vis_curve}) were obtained on a single night and show a similar visibility-baseline trend as the rest of the data, disfavoring variability as an explanation for the observed  trends.  Thus, our conclusion on the presence of smooth gas emission inside the dust destruction radius is robust.
 
\section{Discussion}
In this work, we have conclusively shown the presence of strong NIR emission between the magnetospheric co-rotation radius and  the dust sublimation radius in Herbig Ae stars MWC275 and AB~Aur. The underlying mechanism for this emission is not known requiring high resolution NIR spectroscopy and self-consistent modeling of the gas-dust transition region to identify the gas species and the physical processes responsible for the emission. 

The detection of continuum gas emission within the dust destruction radius through long-baseline interferometry is of special importance to understanding the star-disk connection.  Accretion models for YSOs have tended to focus on two separate physical scales (i) the process of mass and angular momentum transport in the outer disk  and (ii) gas accretion from a few stellar radii on to the stellar photosphere. Milli-meter interferometry and direct imaging in the mid-IR have allowed observational access to the outer disk while high-resolution optical-spectroscopy can probe the circumstellar environment around YSOs on scales of a few stellar radii.  Our near-IR interferometry results on MWC275 and AB~Aur provides exciting prospects for connecting the two scales and determining the structure of the gas flow from a few tenths of an AU down to the stellar surface. In the following paragraph, we describe NIR interferometry as a tool in probing accretion physics inside the dust rim.

In the regime where in the inner gas disk is optically thin, the Shakura-Sunyaev viscosity parameter for the gas disk can be constrained using a combination of  NIR interferometry and optical spectroscopy. NIR interferometry in multiple wavelength bands can constrain the inner gas-disk surface density and temperature profiles. The surface density of the disk is related to the mass accretion rate and the viscosity parameter in quasi-steady-state disks \citep{shakura1973}. Thus, simultaneous measurements of mass in-fall rates (through modeling of optical-emission lines and optical continuum excess above photospheric emission \citep{Calvet1998, muz2001} ) and the gas surface density profiles will  directly constrain the viscosity parameter `$\alpha$' in the inner disk. This is key to  understanding angular momentum transport in circumstellar disks.

\acknowledgements
 We acknowledge contributions from Nuria Calvet, Michael Sitko, PJ Goldfinger, Christopher Farrington and Steve Golden. We also thank the anonymous referee for valuable comments and suggestions. Research at the CHARA Array is supported by the National Science Foundation through
grants AST 06-06958 and AST 03-52723 and by the Georgia State University through the offices of
the Dean of the College of the Arts and Sciences and the Vice President for Research. This project was partially supported by NASA grant 
050283. This publication makes use of NASA's  Astrophysics Data System Abstract Service. The calibrator sizes were obtained with the fBol module of getCal, a package made available by the Michelson  Science Center, California Institute of Technology (http://msc.caltech.edu).

\bibliographystyle{apj}

\bibliography{model_temp}

\clearpage

\begin{table}[h]
\begin{center}
\caption{
MWC275 and AB~Aur model parameters for K-band emission. The models also include extended envelopes contributing 5\% of the K -band emission \citep{monnier2006}.
}
\begin{tabular}{cccccc}
\footnotesize
Star
& model & star & dust-rim  & gas-disk  &rim  \\
 & type&emission  & emission  & emission  & inner radius\\

 &                & (\%)     & (\%)  & (\%)   & (AU)    \\
\hline
MWC275  &dust rim only &25&  69     & -  & 0.19 \\
           &dust rim+gas & 10 &  29     & 56  & 0.22  \\
AB Aur  &dust rim only &22&  73      & -  & 0.21   \\
            &dust rim+gas &12& 45       & 38  & 0.25 \\

\hline
\end{tabular}
\label{modelprops}
\end{center}

\end{table}

\clearpage

%
%



%









\end{document}